\documentclass[sigconf,natbib=true,anonymous=False]{acmart}

\usepackage{subfigure}
\usepackage{bm}
\usepackage{amsmath,amsfonts}
\usepackage{algorithmic}
\usepackage{graphicx}
\usepackage{textcomp}
\usepackage{xcolor}
\usepackage[linesnumbered, ruled]{algorithm2e}
\usepackage{algorithmic}
\usepackage{url}
\usepackage{balance}
\usepackage{booktabs}
\usepackage{makecell}
\usepackage{multirow}
\usepackage{multicol}
\usepackage{hyperref}
\usepackage{proof}
\usepackage{amsmath,amsfonts}

\usepackage{setspace}
\usepackage{amsmath}
\usepackage{pifont}
\newcommand{\cmark}{\ding{51}}%
\newcommand{\xmark}{\ding{55}}%
\usepackage{adjustbox}
\usepackage{makecell}
\usepackage{mathtools}
\usepackage{arydshln}
\AtBeginDocument{%
  }

\copyrightyear{2023}
\acmYear{2023}
\setcopyright{acmlicensed}\acmConference[CIKM '23]{Proceedings of the 32nd
ACM International Conference on Information and Knowledge
Management}{October 21--25, 2023}{Birmingham, United Kingdom}
\acmBooktitle{Proceedings of the 32nd ACM International Conference on
Information and Knowledge Management (CIKM '23), October 21--25, 2023,
Birmingham, United Kingdom}
\acmPrice{15.00}
\acmDOI{10.1145/3583780.3614773}
\acmISBN{979-8-4007-0124-5/23/10}

\begin{document}

\title{AdaMCT: Adaptive Mixture of CNN-Transformer for Sequential Recommendation}

\author{Juyong Jiang}
\authornote{Equal contribution.}
\authornote{Work done when Juyong was AI Global Residency of Upstage.}
\affiliation{
  \institution{The Hong Kong University of Science and Technology (Guangzhou)}
  \streetaddress{38 Zheda Rd}
  \city{}
  \country{}}
\email{csjuyongjiang@gmail.com}

\author{Peiyan Zhang}
\authornotemark[1]
\affiliation{
    \institution{The Hong Kong University of Science and Technology}
    \country{}}
\email{pzhangao@cse.ust.hk}

\author{Yingtao Luo}
\affiliation{%
  \institution{Carnegie Mellon University}
  \country{ }
  }
\email{yingtaoluo@cmu.edu}

\author{Chaozhuo Li}
\authornote{Corresponding authors.}
\affiliation{
    \institution{Microsoft Research Asia}
    \city{}
    \country{}}
\email{cli@microsoft.com}

\author{Jae Boum Kim}
\affiliation{
    \institution{The Hong Kong University of Science and Technology \& Upstage}
    \country{}}
\email{jbkim@cse.ust.hk}

\author{Kai Zhang}
\authornotemark[3]
\affiliation{%
  \institution{East China Normal University}
  \streetaddress{38 Zheda Rd}
  \city{}
  \country{}}
\email{kzhang@cs.ecnu.edu.cn}

\author{Senzhang Wang}
\affiliation{
    \institution{Central South University}
    \country{}}
\email{szwang@csu.edu.cn}

\author{Xing Xie}
\affiliation{
    \institution{Microsoft Research Asia}
    \city{}
    \country{}}
\email{xing.xie@microsoft.com}

\author{Sunghun Kim}
\affiliation{%
  \institution{The Hong Kong University of Science and Technology (Guangzhou)}
  \streetaddress{38 Zheda Rd}
  \city{}
  \country{}}
\email{hunkim@ust.hk}

\renewcommand{\shortauthors}{Jiang and Zhang, et al.}

\begin{abstract}
Sequential recommendation (SR) aims to model users’ dynamic preferences from a series of interactions. A pivotal challenge in user modeling for SR lies in the inherent variability of user preferences. An effective SR model is expected to capture both the long-term and short-term preferences exhibited by users, wherein the former can offer a comprehensive understanding of stable interests that impact the latter. To more effectively capture such information, we incorporate locality inductive bias into the Transformer by amalgamating its global attention mechanism with a local convolutional filter, and adaptively ascertain the mixing importance on a personalized basis through layer-aware adaptive mixture units, termed as AdaMCT. Moreover, as users may repeatedly browse potential purchases, it is expected to consider multiple relevant items concurrently in long-/short-term preferences modeling. Given that softmax-based attention may promote unimodal activation, we propose the Squeeze-Excitation Attention (with sigmoid activation) into SR models to capture multiple pertinent items (keys) simultaneously. Extensive experiments on three widely employed benchmarks substantiate the effectiveness and efficiency of our proposed approach. Source code is available at \href{https://github.com/juyongjiang/AdaMCT}{\color{blue}{https://github.com/juyongjiang/AdaMCT}}.
\end{abstract}




\maketitle

\section{Introduction}
Recommender systems facilitate the provision of tailored information to users, taking into account their preferences as evidenced by historical interactions~\citep{guo2022evolutionary}, which are widely applied in e-commerce websites, web searches, and so forth~\citep{fang2020deep,zhang2019deep}. Analogous with words of sentences, a user’s interactions with items naturally constitute a behavioral sequence. Thus, sequential recommendation (SR) has been proposed to encapsulate the sequential characteristics of user behaviors, thereby modeling users' preferences more effectively~\citep{hidasi2015session,ji2020sequential}.

\begin{figure}[t]
\centering
\includegraphics[width=0.8\linewidth]{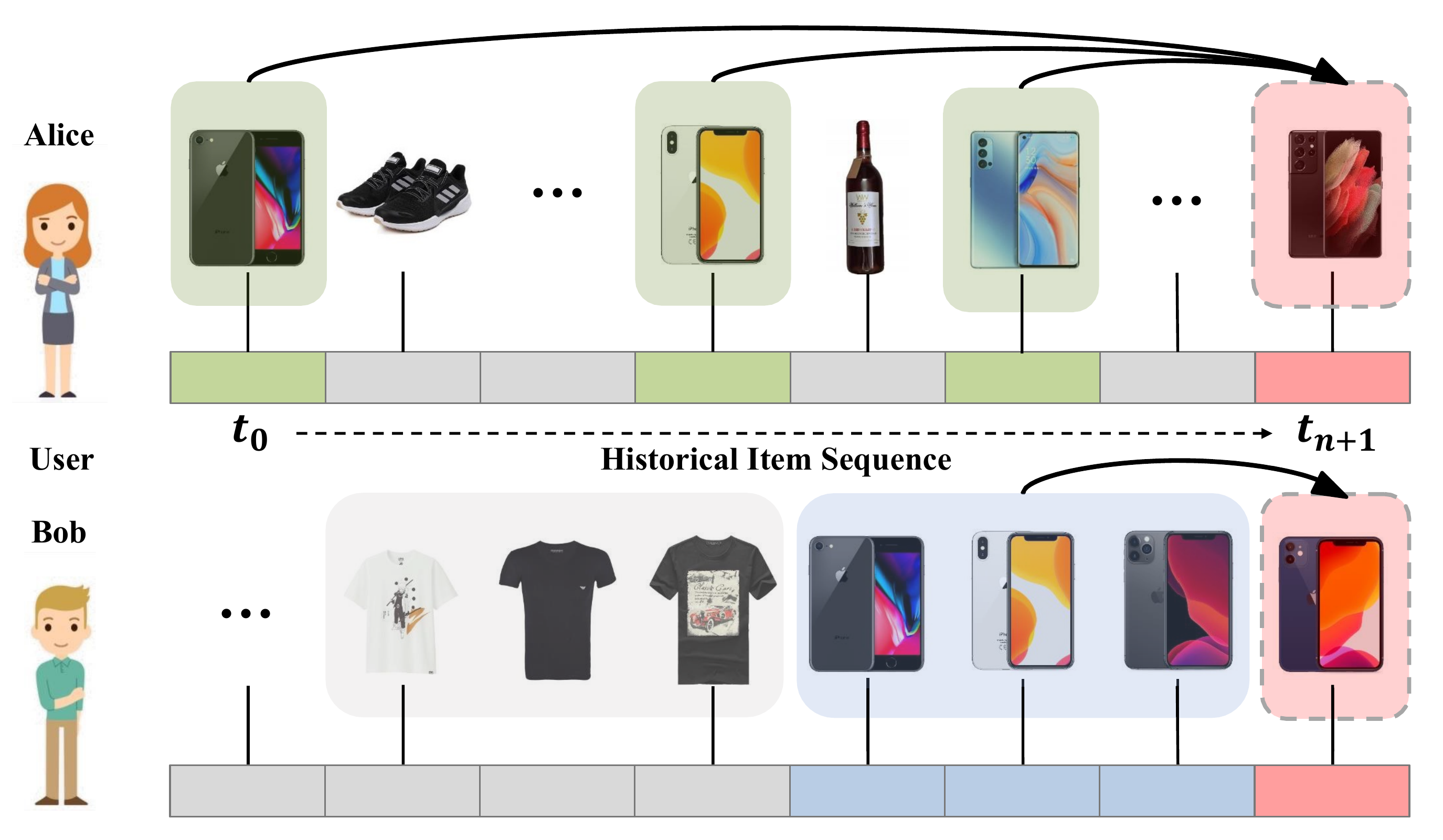}
\caption{Personalized global interactions and local features dependency. Alice and Bob have different interest patterns. For Alice, the next item relies more on her global interactions interest, while for Bob, he shows a local features dependency on his previous historical item sequences.}
\label{fig:fig1}
\end{figure}

A vital challenge of user modeling for SR lies in the inherent variability of user preferences. In large-scale recommender systems, millions of new items are generated and incorporated into the candidate pool on a daily basis~\citep{zhang2023survey}. It is imperative for recommender systems to promptly and accurately discern users' short-term preferences, as their attention is readily captivated by emerging trends and popular topics. In contrast, users' long-term preferences offer a valuable insight into their aggregated, stable interests, effectively complementing their short-term interests. Hence, it is crucial for SR models to adeptly capture both the dynamic nature of short-term preferences and the enduring stability of long-term preferences exhibited by users.

Fortunately, recent studies in SR have endeavored on this target by modeling and integrating long- and short-term preferences prior to making recommendations. The initial line of research~\citep{vaswani2017attention, dong2021attention,zhang2023efficiently} concentrates on capturing long-term preferences while employing the most recently interacted item as a representation of short-term preferences. However, due to the SR-related inductive bias, specifically local invariance~\citep{chen2019session}, which suggests that the local order of the last few items is inconsequential, it becomes arduous to comprehensively represent a user's short-term interests using only the final item. Furthermore, these works merely fuse the long-/short-term preferences through summation or concatenation~\citep{hu2018local, xu2019recurrent, chen2019session, xu2020joint, he2020consistency, liu2020deep, yu2020tagnn}. As a result, these approaches fail to obtain personalized weights for long-/short-term preferences pertaining to different items for each user.

Another line of works adopts a parallel paradigm to model the long-/short-term preferences respectively with recurrent and attentive models~\citep{hu2017diversifying,liu2018stamp,zhou2019deep} or even explicit user behavior graphs~\citep{xie2022long}. Subsequently, they implement a gating mechanism~\citep{gu2020improving} to facilitate the integration of personalized long-term and short-term preferences. Nevertheless, the long-/short-term preference modeling processes are segregated. Given that users' long-term preferences can offer a comprehensive understanding of stable interests that influence short-term interests and vice versa~\citep{yu2019multi,an2019neural,devooght2017long}, this paradigm may adversely affect the comprehension of users' overall preferences. Therefore, it is reasonable that the long-/short-term preferences should be mutually reinforced in the context of user modeling.

In particular, modeling the long-/short-term preferences in SR is non-trival, primarily encountering the following three challenges: (1) \textit{How can long-term and short-term preferences be modeled conjointly?} As users’ long-term preferences could provide users’ aggregated stable interests that influence the short-term interests and vice versa~\citep{xie2022long}, these preferences should engage in a mutual reinforcement within user modeling. (2) \textit{How to integrate the short-term and long- term preferences effectively?} Both types of interests play a crucial role in real-world recommendations, despite the inherent biases between them. Models must be capable of discerning which interest is more pertinent in varying situations. Previous studies~\citep{hu2018local, xu2019recurrent, chen2019session,xie2022long} have relied on simple summation, concatenation, or a singular gating mechanism, leading to restricted model capacity. (3) \textit{How can multiple relevant items be considered concurrently in long-term and short-term preference modeling?} As illustrated in Figure~\ref{fig:fig1}, users may repeatedly browse and compare potential purchases, resulting in more than one item being highly correlated with their preferences. However, the softmax activation in the attention scheme necessitates that the sum of all item scores equals 1, creating mutually exclusive relations~\citep{hu2018squeeze} that impede the simultaneous activation of multiple highly-related items. These three challenges are of significant importance in real-world scenarios, yet there is a scarcity of research that systematically addresses them in unison.

To address these challenges, we introduce a novel Adaptive Mixture of CNN-Transformer for Sequential Recommendation (AdaMCT) that explicitly learns personalized behavior preferences to adaptively leverage locality inductive bias (CNNs)~\citep{Caser} and global interactions (Transformer)~\citep{kang2018self, sun2019bert4rec} modeling. Specifically, we employ a local-global dependencies mechanism comprising both (local) convolutional layers and (global) self-attention layers, facilitating the joint modeling of long-term and short-term preferences. The CNN layer exhibits heightened sensitivity to short-term interest patterns, while the self-attention layer is better suited for identifying long-term dependency patterns. To effectively integrate long-term and short-term preferences, we devise adaptive mixture units that decouple the fusion process into multiple layers, thereby enhancing expressibility and adaptively aggregating long-term and short-term preferences for personalized user modeling. Furthermore, we propose the Squeeze-Excitation Attention to generate attention scores for both short-term and long-term information branches. Squeeze-Excitation Attention eliminates the sum-up-to-1 constraint, allowing for the consideration of multiple relevant purchases simultaneously and enhancing model expressibility. We validate the AdaMCT approach on three public benchmarks. Substantial improvements across multiple indicators demonstrate the efficacy of the proposed methods. The main contributions of this paper are as follows:
\begin{itemize}
    \item We incorporate the locality inductive bias (CNNs) into the Transformer structure to model the long-/short-term preferences conjointly, and devise the adaptive mixture
units that decouple the fusion process into multiple layers, thereby enhancing expressibility and adaptively aggregating long-term and short-term preferences for personalized user modeling.
    \item We propose Squeeze-Excitation Attention (SEAtt) to replace the softmax operation in both local and global information branches to simultaneously take into account multiple relevant purchases to effectively improve the representation learning of sequential dependencies.
    \item Extensive experiments on three public datasets demonstrate that our proposed solution achieves comparable (or better) performance than the previous Transformer and CNNs-based models. Ablation studies show the efficacy of our proposed components.
\end{itemize}

\section{Related Work}
\subsection{Sequential Recommendation}
Sequential recommendation aims to predict the future item or item list based on a historical user behavior sequence. In the literature, early works mainly use Markov Chain models \citep{koren2009matrix, rendle2010factorization, he2016fusing} that make personalized next recommendations conditioned on the previous one's behavior. These methods only capture adjacent local sequential patterns. Later on, the genre of neural network-based models starts to rise. Recurrent Neural Networks (RNNs) improve the recommendation performance in many tasks with sequential dependencies and memory mechanisms \citep{liu2016predicting, liu2016context, yu2016dynamic, cui2018mv}. RNNs impose strict sequential order in representation learning and face the global dependency problem when the sequence is too long. Meanwhile, Convolutional Neural Networks (CNNs) are used to capture local features across the sequence, emphasizing the impact of more recent behaviors \citep{Caser, yuan2019simple}. More recently, attentive modules are extensively studied in the sequential recommendation venue, resulting in extraordinary performances \citep{kang2018self, kang2019recommender}. Self-attention models allow point-to-point feature interaction within the item sequence, which solves the global dependency problem and uses a longer sequence with more information \citep{kang2018self, luo2021stan, jiang2021sequential}. Meanwhile, graph neural networks have been attempted to discover the graph-based topological interaction of sequences \citep{wu2019session, ma2020memory, wang2020next}. Transformer and Bidirectional Transformer (BERT) that combine multiple blocks including attention layers, feed-forward networks, and position embeddings also report state-of-the-art performances in sequential recommendation tasks \citep{chen2019behavior, sun2019bert4rec, wu2020sse}. Mixed models \citep{yu2019adaptive, xu2019recurrent, chen2019session} that consider both global and local preferences have also received extensive attention in recent years. 

\subsection{Local-Global Awareness}
Local-global awareness is a topic that has raised discussions in the venue of recommendation systems.
Many prior works \citep{song2020local, huang2020meta, ma2019hierarchical} propose and design local-global memory mechanisms to learn different-level item sequences, while they use a simple concatenation of neural modules without an adaptive learnable preference factor. 
However, some works \citep{he2020consistency, lin2020fissa} also consider the adaptive mixture of local-global representation learning and show some advantages, but the difference from previous works is that we first attempt to adopt CNNs and Transformer to capture local and global dependencies in SR.  Moreover, Local-global regularization is used to learn the local and global item correlations with both deep neural network and matrix factorization \citep{zhang2016global, liu2020collaborative}. These works design user-specific strategies to fuse local and global information into the recommendation model \citep{christakopoulou2016local, tan2021sparse}. Other works focus on improving neural architectures. Global and local sparse linear method models are combined to make recommendations for different user preferences \citep{hu2018local}. A recurrent convolutional network is proposed to learn complex global sequential dependency with RNNs and extract local dependency using CNNs \citep{xu2019recurrent}, which marks a combined use of sequence order bias and locality bias. An encoder for modeling the global cross-session dependency is proposed to improve the self-attention model \citep{xu2020joint}. Local generative models that capture different user preferences are combined to make a global prediction \citep{liu2020deep}. Local invariance is considered in session-based recommendation \citep{chen2019session} as extra local information to substitute the main model. Attention and graph network are combined to model different user interests \citep{yu2020tagnn} with graph topology inductive bias. Recently, some papers \citep{yu2019adaptive, luo2020metaselector} find that the personalized and adaptive use of model components can effectively improve the model performance on RNNs. However, most works combining multiple neural modules, especially the state-of-the-art Transformer, still rarely consider the user-specific learnable adaptation. Moreover, they hardly consider the comprehensive inductive biases from locality to global dependency, and from sequential order to positional equivalence.

\begin{figure*}
\centering
\includegraphics[width=.8\linewidth]{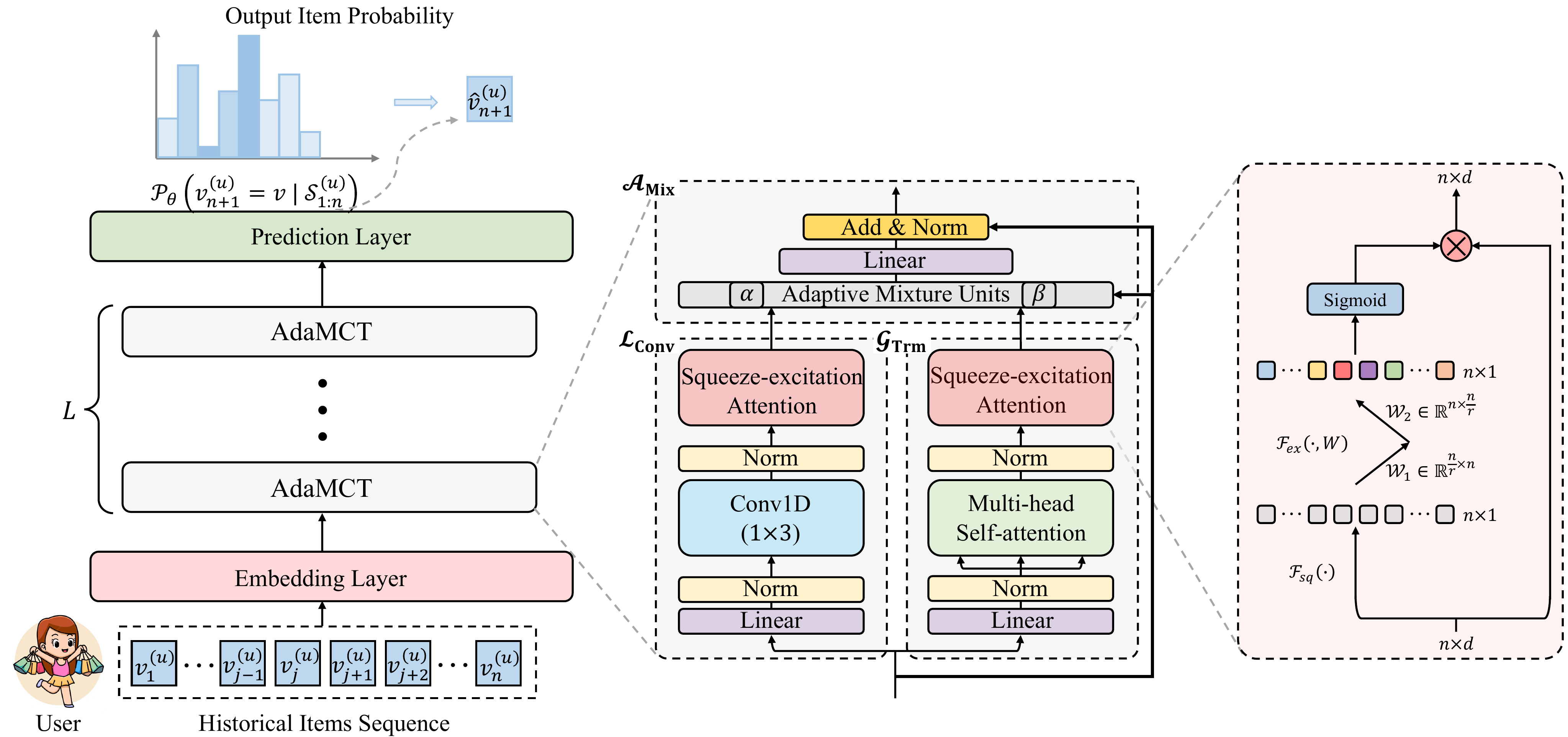}
\caption{The model architecture of the proposed AdaMCT. AdaMCT first generates item embedding with positional embedding by embedding layer, then aggregates sequence representations through $L\times$ AdaMCT layers, in which each layer includes three sub-layers, \emph{i.e.}, global attention module ($\mathcal{G}_\text{Trm}$), local convolutional module ($\mathcal{L}_\text{Conv}$), and adaptive mixture units ($\mathcal{A}_\text{Mix}$). Finally, a prediction layer generates a recommendation score for each candidate item.}
\label{fig:fig2}
\end{figure*}

\section{Methodology}

\subsection{Problem Statement}
$\mathcal{U}=\{u_1, u_2, ..., u_{|\mathcal U|}\}$ denotes users, $\mathcal{V}=\{v_1, v_2, ..., v_{|\mathcal{V}|}\}$ denotes items, and $\mathcal{S}^{(u)}_{1:n}=[v_1^{(u)}, v_2^{(u)}, ..., v_n^{(u)}]$ denotes the interaction sequence in chronological order of user $u \in \mathcal{U}$. Each $v_i^{(u)} \in \mathcal{V}$ refers to an item that $u$ has interacted with at time step $i$. Sequential recommendation aims to predict the item that user $u$ will interact with at the $(n+1)$th timestamp. The conditional probability of each item $v \in \mathcal{V}$ becoming the $(n+1)$th item is denoted as
$\mathcal{P}_\theta(v_{n+1}^{(u)}=v| \mathcal{S}^{(u)}_{1:n})$, where $\theta$ represents the model parameters.


\subsection{Model Architecture}
The overall model architecture is shown in Figure \ref{fig:fig2}. Our proposed AdaMCT consists of 1) an embedding module that learns the dense representations of items, including item embeddings and positional embeddings; 2) a multi-stacked Adaptive Mixture of CNN-Transformer blocks, in which both comprise a global attention module (Transformer) that learns the global user interest preferences with squeeze-excitation attention (SEAtt), a local convolutional module (CNNs) that learns user local interest pattern with SEAtt, and a module- and layer-aware adaptive mixture units, that makes a personalized balanced use of local and global dependencies modules; and 3) an output layer with cross-entropy loss that computes the matching probability.

\subsection{Embedding Module}
The embedding module contains an item embedding layer and a positional embedding layer. The item embedding layer ${E}_{item}\in \mathbb{R}^{|\bm{\mathcal V}|\times d_{model}}$ with the latent dimension of $d_{model}$ encodes each scalar item ID $v_i^{(u)} \in \mathbb{R}$ into a dense latent vector representation $e_i^{(u)} \in \mathbb{R}^{d_{model}}$, aiming to reduce computation and improve representation learning. The positional embeddings $\mathcal{P}os\in \mathbb{R}^{n\times d_{model}}$ we follow  \cite{kang2018self} to employ learnable embeddings and maintain the same embedding dimensions of the item embedding. The overall embedding module transforms the sequence of $\mathcal{S}^{(u)}_{1:n}=[v_1^{(u)}, v_2^{(u)}, ..., v_n^{(u)}] \in \mathbb{R}^{n}$ into ${E}_{item}^{pos}(\mathcal{S}^{(u)}_{1:n})= [{E}_{item}(\mathcal{S}^{(u)}_{1:n}) + \mathcal{P}os(\mathcal{S}^{(u)}_{1:n})]\in \mathbb{R}^{n \times d_{model}}$.

\subsection{Adaptive Mixture of CNN-Transformer}
AdaMCT has three sub-layers, a global attention module (Transformer) with SEAtt referred to as $\mathcal{G}_\text{Trm}$, a local convolutional module (CNNs) with SEAtt referred to as $\mathcal{L}_\text{Conv}$, and adaptive mixture units referred to as $\mathcal{A}_\text{Mix}$, as shown in the middle of Figure \ref{fig:fig2}. 

\subsubsection{Global Attention Module} 
As shown in Figure \ref{fig:fig2}, we first utilize a linear layer, followed by layer normalization \citep{ba2016layer}, to adjust latent dimensionality and enhance the non-linear representation. Then, a multi-head self-attention layer (MHSA) followed by layer normalization is adopted to capture global dependencies of sequence. 
Finally, a squeeze-excitation attention layer (SEAtt) to re-weight the importance of each item from the sequence level. Formally, 
\begin{equation}
\begin{aligned}
\mathcal{G}_\text{Trm}^l=\mathrm{SEAtt}(\mathrm{MHSA}(\mathcal{E}^{en}_{global}(\mathcal{H}_{(u)}^{l-1})) \otimes \text{MHSA}(\mathcal{E}^{en}_{global}(\mathcal{H}^{l-1}_{(u)}))),
\end{aligned}
\end{equation}
\begin{align}
\text{MHSA}(\mathcal{E}^{en}_{global}(\mathcal{H}_{(u)}^{l-1}))=\text{LN}(\text{Dpt}(\mathcal{C}onat(\text{head}_1, ...,\text{head}_h)\mathcal{W}^\mathcal{O})),
\end{align}
\begin{align}
\text{head}_i = \text{Att}(\underbrace{\mathcal{E}^{en}_{global}(\mathcal{H}_{(u)}^{l-1})\mathcal{W}_i^\mathcal{Q}}_{\mathcal{Q}},\underbrace{\mathcal{E}^{en}_{global}(\mathcal{H}_{(u)}^{l-1})\mathcal{W}_i^\mathcal{K}}_{\mathcal{K}}, \nonumber \\ \underbrace{\mathcal{E}^{en}_{global}(\mathcal{H}_{(u)}^{l-1})\mathcal{W}_i^\mathcal{V}}_\mathcal{V}),
\end{align}
\begin{align}
\text{Att}(\mathcal{Q},\mathcal{K},\mathcal{V})=\mathrm{softmax}(\frac{\mathcal{Q}\mathcal{K}^\mathcal{T}}{\sqrt{d_{model}/h}} + \mathcal{M}_{mask})\mathcal{V},
\end{align}
\begin{align}
\mathcal{E}^{en}_{global}(\mathcal{H}_{(u)}^{l-1}) = \text{LN}(\text{Dpt}(\mathcal{L}inear_{en}(\mathcal{H}_{(u)}^{l-1}))), \mathcal{H}_{(u)}^{0}=\mathcal{E}(\mathcal{S}^{(u)}_{1:n}).
\end{align}
\begin{equation}
\begin{aligned}
    \mathcal{M}_{mask} = [a_{ij}] = 
    \begin{cases}
    a_{ij} = -\infty& \text{for $i < j$ } \\
    a_{ij} = 0 & \text{otherwise}
    \end{cases}
\end{aligned}
\end{equation}
where $\mathcal{G}_\text{Trm} \in \mathbb{R}^{n \times d_{model}}$ is the output of the $l$-th global attention module, $\text{SEAtt} (\cdot) \in \mathbb{R}^{n}$ denotes sequence-level attention scores. The details of SEAtt module will be discussed in the following section. ``$\otimes$'' denotes the element-wise operation with the broadcasting mechanism. $\mathcal{H}_{(u)}^{l-1} \in \mathbb{R}^{n\times d_{model}}$ represents the $(l-1)$-th AdaMCT layer's output, LN and Dpt denote layer normalization and dropout operations, respectively. $h$ represents the number of attention layers. $\mathcal{W}_i^\mathcal{Q} \in \mathbb{R}^{d_{model} \times d_{model}/ h}$, $\mathcal{W}_i^\mathcal{K} \in \mathbb{R}^{d_{model} \times d_{model}/ h}$, $\mathcal{W}_i^\mathcal{V} \in \mathbb{R}^{d_{model} \times d_{model} / h}$, and $\mathcal{W}_i^\mathcal{O} \in \mathbb{R}^{d_{model} \times d_{model}/h}$ are the parameters that are unique per layer and attention head and used to transform the Query $\mathcal{Q}$, Key $\mathcal{K}$, Value $\mathcal{V}$, and the output of MHSA layer. The effect of ``temperature'' $\sqrt{d_{model}/h}$ is to avoid overly large values of the inner product and extremely small gradients to produce a softer attention distribution \citep{vaswani2017attention,kang2018self,sun2019bert4rec}. 
For a casual attention mask $\mathcal{M}_{mask}$, the upper triangular part of it is set to $-\infty$ otherwise 0, representing each item only attends to itself and preceding items.   

\subsubsection{Local Convolutional Module} This module is shown in the left part of the AdaMCT in the middle of Figure \ref{fig:fig2}. Similar to the global attention module ($\mathcal{G}_\text{Trm}$), each local convolutional module ($\mathcal{L}_\text{Conv}$) first utilize a linear layer, followed by layer normalization \citep{ba2016layer}, to adjust latent dimensionality and enhance the non-linear representation. Formally,
\begin{align}
\mathcal{E}^{en}_{local}(\mathcal{H}_{(u)}^{l-1}) = \text{LN}(\text{Dpt}(\mathcal{L}inear_{en}(\mathcal{H}_{(u)}^{l-1}))), \mathcal{H}_{(u)}^{0}=\mathcal{E}(\mathcal{S}^{(u)}_{1:n}),
\end{align}

Subsequently, we adopt a 1D convolutional layer with $k=3$ kernel size followed by layer normalization on $\mathcal{E}^{en}_{local}(\mathcal{H}_{(u)}^{l-1}) \in \mathbb{R}^{n \times d_{model}}$. Formally, considering $m$ 1D convolutional filters $\mathcal{F}^i \in \mathbb{R}^{k \times d_{model}}$, $i \in \{1,2,...,m\}$ with $k$ kernal size, $1\le k \le n$. $\mathcal{F}^i$ will slide from left to right to interact $\mathcal{E}^{en}_{local}(\mathcal{H}_{(u)}^{l-1})$ with every successive $k$ item along the size (length)-axis. Notably, we use the zero paddings and stride to 1 to maintain the size (length) dimension invariant in the convolutional layer. Thus, after sliding in all the positions $j \in \{1,2,...,n\}$, the final output of 1D convolutional layer can be formalized by 
\begin{align}
&\text{Conv}^{k=3}_{mid}(\mathcal{E}^{en}_{local}(\mathcal{H}_{(u)}^{l-1}))=\text{LN}(\text{Dpt}([\text{Conv}^{k=3}_{mid}(\mathcal{E}^{en}_{local}(\mathcal{H}_{(u)}^{l-1}))_1, \nonumber\\&..., \text{Conv}^{k=3}_{mid}(\mathcal{E}^{en}_{local}(\mathcal{H}_{(u)}^{l-1}))_j,..., \text{Conv}^{k=3}_{mid}(\mathcal{E}^{en}_{local}(\mathcal{H}_{(u)}^{l-1}))_n]))
\end{align}
\begin{align}
\text{Conv}^{k=3}_{mid}(\mathcal{E}^{en}_{local}(\mathcal{H}_{(u)}^{l-1}))_j=\mathcal{C}oncat(\mathcal{C}_j^1, ..., \mathcal{C}_j^m) \in \mathbb{R}^{m},
\end{align}
\begin{align}
\mathcal{C}_j^i=\Phi (\mathcal{E}^{en}_{local}(\mathcal{H}_{(u)}^{l-1})_{[j:j+k-1]}\mathcal{F}^i),
\end{align}
where $\Phi(\cdot)$ denotes the activation function of convolutional layers in which we use the $ReLU$ by default. $\mathcal{E}^{en}_{local}(\mathcal{H}_{(u)}^{l-1})_{[j:j+k-1]} \in \mathbb{R}^{k \times d_{model}}$ represents the sub-sequence with indexes from $j$ to $j+k-1$. 
$\text{Conv}^{k=3}_{mid}(\mathcal{E}^{en}_{local}(\mathcal{H}_{(u)}^{l-1}))_j \in \mathbb{R}^{m}$ denotes the convolutional value in the position $j$.

Furthermore, we input the above output of convolutional layer to the SEAtt module to re-weight the importance of each item from the sequence level. Formally, 
\begin{align}
\mathcal{L}_\text{Conv}=\text{SEAtt}(\text{Conv}^{k=3}_{mid}(\mathcal{E}^{en}_{local}(\mathcal{H}_{(u)}^{l-1}))) \otimes \text{Conv}^{k=3}_{mid}(\mathcal{E}^{en}_{local}(\mathcal{H}_{(u)}^{l-1})).
\end{align}
where $\mathcal{L}_\text{Conv} \in \mathbb{R}^{n \times m}$ represents the output of local convolutional module. In this work, we set $m=d_{model}$ to make $\text{L}_\text{Conv}$ have the same dimension as $\mathcal{G}_\text{Trm}$, \emph{i.e.}, $\mathcal{L}_\text{Conv} \in \mathbb{R}^{n \times d_{model}}$.
\begin{figure*}
    \centering
    \includegraphics[width=0.8\linewidth]{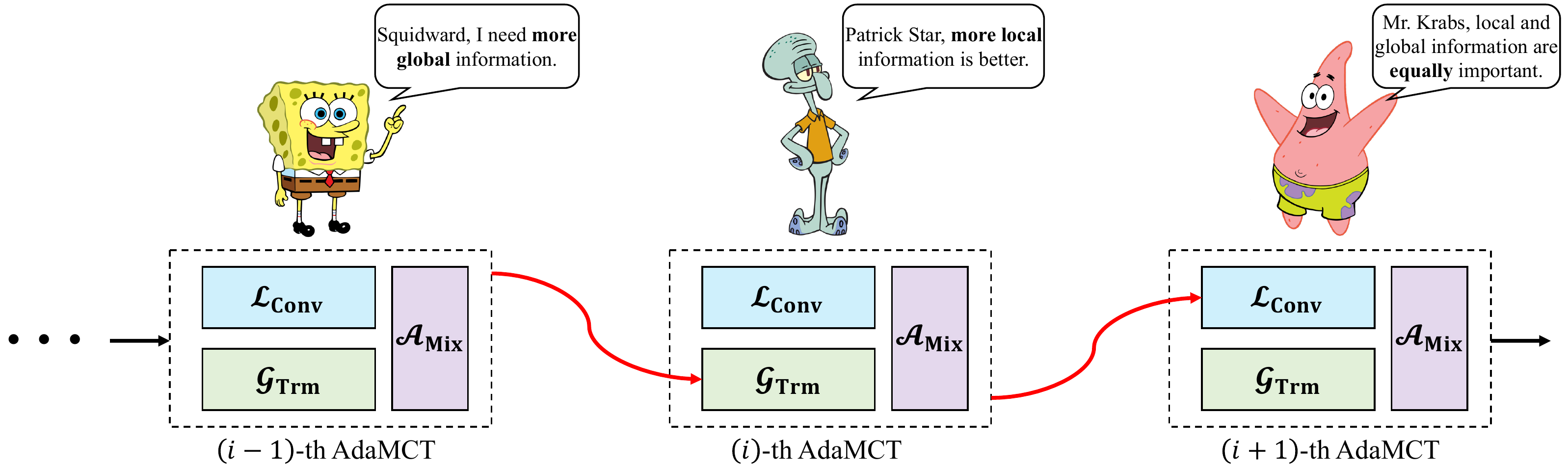}
    \caption{A toy example to show how the model self-adaptively coordinates the mixture of local (CNNs) and global (Transformer) information between different layers. The cartoon characters are from an American animated television series SpongeBob SquarePants.}
    \label{fig:toy_example}
\end{figure*}
\subsubsection{Squeeze-Excitation Attention} In practical scenarios, users' behaviors are dynamic over time which leads to more than one historical user-item interaction having a high correlation to the next item. However, many previous works \citep{guo2019attentive,dong2021attention} neglect the impact of \emph{multi-high-related} user-item interactions on the next item, which results in sub-optimal recommendation performance. As can be seen in Figure \ref{fig:fig1}, Alice and Bob have \emph{multiple} historical items \emph{iPhone} which are high-related to the next items \emph{iPhone}. To address this issue, inspired by \citep{hu2018squeeze} which recalibrates channel-wise feature by explicitly modelling inter-dependencies between channels, we propose Squeeze-Excitation Attention. To adapt to the recommendation scenario, we extend it to propose an \emph{item-wise} statistics. Formally, the attention scores can be yielded by  
\begin{align}
    \text{SEAtt}(\text{Mod}(\mathcal{H}_{(u)}^{l-1}))=\mathcal{F}_{ex}(\mathcal{Z}^{(u)}),
\end{align}
\begin{align}
    \mathcal{F}_{ex}(\mathcal{Z}^{(u)})=\sigma(g(\mathcal{Z}^{(u)}))=\sigma(\mathcal{W}_2\delta(\mathcal{W}_1\mathcal{Z}^{(u)})),
\end{align}
\begin{align}
    \mathcal{Z}^{(u)}=\mathcal{F}_{sq}(\text{Mod}(\mathcal{H}_{(u)}^{l-1}))=\frac{1}{d_{model}}\sum_{j=1}^{d_{model}}{\text{Mod}(\mathcal{H}_{(u)}^{l-1})}_{[:,j]}.
\end{align}
where $\text{SEAtt}(\text{Mod}(\mathcal{H}_{(u)}^{l-1})) \in \mathbb{R}^{n \times 1}$ denotes the attention scores. $\mathcal{Z}^{(u)} \in \mathbb{R}^{n \times 1}$ denotes \emph{item-wise} statistics produced by average pooling of $\text{Mod}(\mathcal{H}_{(u)}^{l-1}) \in \mathbb{R}^{n \times d}$ which denotes the input of SEAtt in $\mathcal{G}_\text{Trm}$ or $\mathcal{L}_\text{Conv}$, \emph{i.e.}, $\text{MHSA}(\mathcal{E}^{en}_{global}(\mathcal{H}_{(u)}^{l-1})))$ or $\text{Conv}^{k=3}_{mid}(\mathcal{E}^{en}_{local}(\mathcal{H}_{(u)}^{l-1}))$. $\mathcal{W}_1 \in \mathbb{R}^{\frac{n}{r} \times n}$ and $\mathcal{W}_2 \in \mathbb{R}^{n \times \frac{n}{r}}$ denote squeeze and excitation parameters, respectively. The reduction ratio $r$ is a hyper-parameter with a default value $2$ to limit model complexity and benefit generalization. $\delta$ and $\sigma$ refer to the \emph{ReLU} and \emph{Sigmoid} activation function, respectively.  

\textbf{\emph{Discussion}} Squeeze-Excitation Attention (SEAtt) differs from the widely used attention \cite{kang2019recommender,ji2020sequential,dong2021attention} in two aspects: \textbf{a)} adopting effective \emph{squeeze-excitation} operation to improve the representation capability \cite{hu2018squeeze}; \textbf{b)} employing the \emph{non-mutually-exclusive} relations for recommendation which allows the next item to interact with multiple historical items. In contrast, the $\text{softmax}$ activation requires the sum of all item scores to be 1, which are \emph{mutually exclusive} relations that hinder multiple high-related items to be activated simultaneously. 
\subsubsection{Adaptive Mixture Units} This module is shown in the top part of the AdaMCT in the middle of Figure \ref{fig:fig2}. The adaptive mixture units adjust the personalized trade-off between the global attention module $\mathcal{G}_{\text{Trm}}$ and the local convolutional module $\mathcal{L}_{\text{Conv}}$. Specifically, it first leverages the average pooling operator to produce dimension-wise statistics, and then a linear layer with $\Omega$ activation function is employed to get the adaptive coefficient for each user.
\begin{align}
    \text{Ada}^{(u)}(\mathcal{H}_{(u)}^{l-1})=\Omega(\mathcal{L}inear_{mid}(\mathcal{P}ool(\mathcal{H}_{(u)}^{l-1}))),
\end{align}
\begin{align}
    \mathcal{P}ool(\text{Out}(\mathcal{H}_{(u)}^{l-1})))=\frac{1}{n}\sum_{i=1}^{n}\text{Out}(\mathcal{H}_{(u)}^{l-1})_i,
\end{align}
where $\mathcal{P}ool(\mathcal{H}_{(u)}^{l-1})) \in \mathbb{R}^{d_{model}}$ denotes user preference representation at $(l-1)$-th layer. $\Omega(\cdot)$ denote the $Sigmoid$ activation function which is used to scale the value to the range 0 to 1. $\mathcal{H}_{(u)}^{l-1}$ represents the output of $(l-1)$-th layer. $\text{Ada}^{(u)}(\mathcal{H}_{(u)}^{l-1}) \in \mathbb{R}$ is a personalized coefficient changing over different users since user preference representation is formulated by the average pooling of historical user behaviors. Thus, if we have $|\mathcal U|$ users, it will produce $|\mathcal U|$ coefficient values. 
Besides, the adaptive coefficient values are unique per the AdaMCT layer, which further extends the capacity of representation learning. In all, adaptive mixture units are module- and layer-aware with personalized inductive bias.

Finally, we mix the output of global attention module ($\mathcal{G}_\text{Trm}$) and local convolutional module ($\mathcal{L}_\text{Conv}$) with the adaptive coefficient ($\text{Ada}_{(u)}$) to produce local-global dependencies representation.
\begin{align}
    \mathcal{{M}}^\text{Conv}_\text{Trm}=\text{Ada}^{(u)}(\mathcal{H}_{(u)}^{l-1}) \otimes \mathcal{L}_\text{Conv}+(1-\text{Ada}^{(u)}(\mathcal{H}_{(u)}^{l-1})) \otimes \mathcal{G}_\text{Trm}
\end{align}
where $\text{Ada}^{(u)}(\mathcal{H}_{(u)}^{l-1})$ and $1-\text{Ada}^{(u)}(\mathcal{H}_{(u)}^{l-1})$ are denoted by $\alpha$ and $\beta$ respectively in Figure \ref{fig:fig2}. Furthermore, to maintain the dimension invariant, we apply a linear layer to adjust the latent dimension of $\mathcal{M}^\text{Conv}_\text{Trm}$ and leverage a residual connection \cite{he2016deep} around the input to make the $(l-1)$-th layer's output $\mathcal{H}_{(u)}^{l} \in \mathbb{R}^{n \times d}$ followed by layer normalization. 
\begin{align}
\mathcal{H}_{(u)}^{l}= \text{LN}(\mathcal{H}_{(u)}^{l-1}+\text{Dpt}(\mathcal{L}inear_{out}(\mathcal{{M}}^\text{Conv}_\text{Trm}))).
\end{align}
\subsection{Output Layer}
After $L$ layers hierarchically aggregate information across all positions, we get the final output $\mathcal{H}_{(u)}^L \in \mathbb{R}^{n \times d_{model}}$ for all items of the input sequence. Then, we use the contextual hidden representation at the $n$-th time step of $\mathcal{H}^L_{(u)}[n]$ as the input of the prediction layer, which consists of a two-layer feed-forward network and a $\text{softmax}$ function to produce the \emph{full-ranked} items probability distribution over the next items.
\begin{align}
    \mathcal{P}_{out}(\hat{v}_{n+1}^{(u)}) = \sigma(\delta(\mathcal{H}^L_{(u)}[n]\mathcal{W}^{Pred}+b^{Pred})E_{table}^\mathcal{T}+b^\mathcal{O}),
\end{align}
where $\mathcal{W}^{Pred} \in \mathbb{R}^{d_{model} \times d_{model}}$, $b^{Pred} \in \mathbb{R}^{d_{model}}$, and $b^\mathcal{O} \in \mathbb{R}^{|\mathcal{V}|}$ are parameters and $E_{table}^T \in \mathbb{R}^{d \times |\mathcal{V}|}$ is an item embedding table. We follow \cite{kang2018self, sun2019bert4rec} to use a shared item embedding table of the embedding module to avoid overfitting and speed up model convergence. $\delta$ and $\sigma$ denote the $GELU$ and $softmax$ function, respectively. 

\subsubsection{Model Training} 
In this work, we adopt cross-entropy loss to supervised model training. Formally, the objective loss is formulated as follows:
\begin{align}
    \mathcal{L}_{ce}(\theta) = -\sum_{u \in \bm{\mathcal U}}\log\mathcal{P}_\theta(v_{n+1}^{(u)}=v| \mathcal{S}^{(u)}_{1:n}) = -\sum_{u \in \bm{\mathcal U}}\log\mathcal{P}_{out}(\hat{v}_{n+1}^{(u)}) .
\end{align}
where $\theta$ represents the model parameters.


\section{Experiments}
In this section, we present the empirical evaluations AdaMCT on three widely used benchmarks with varied domains and statistics, aiming to answer the following research questions.
\begin{itemize}
\item \textbf{RQ1:} Can our proposed AdaMCT outperform the state-of-the-art baselines for sequential recommendation?
    \item \textbf{RQ2:} What is the influence of various component designs in AdaMCT?
    \item \textbf{RQ3:} How do different hyper-parameters affect AdaMCT?
    \item \textbf{RQ4:} How efficient is AdaMCT compared to the baselines? How about the model complexity of AdaMCT?
    \item \textbf{RQ5:} Does the visualization of attention matrices of Transformer provide some interpretability for feature representations in the AdaMCT framework?
\end{itemize}

\subsection{Experimental Settings}
\subsubsection{Dataset} 
In this work, our experiments are conducted on three real-world and widely used benchmarks. 
\begin{itemize}
    \item \textbf{Amazon Toys, Sports, Beauty\footnote{\scriptsize{\url{http://snap.stanford.edu/data/amazon/}}}:} This is a series of datasets \citep{mcauley2015image} that consist of large corpora of product reviews crawled from Amazon.com. The datasets are split by the top-level product categories. We adopt the ``\emph{Toys and Games}'', ``\emph{Sports and Outdoors}'', and ``\emph{Beauty}'' categories.
\end{itemize}

The selected datasets are the ones used in important prior works \citep{he2017translation,kang2018self, li2020time,sun2019bert4rec,cho2020meantime}, from which we followed the same data preprocessing procedures. In essence, we convert each dataset into an implicit dataset by treating each rating or review as a presence of a user-item interaction. After that, we group the interactions by user IDs and sort them by timestamps to form a sequence for each user. We follow the same practice as in \citep{he2017neural,kang2018self,rendle2010factorizing,tang2018personalized,sun2019bert4rec,cho2020meantime} to discard short sequences less than five interactions. Table \ref{tab:tab1} describes the statistics of the datasets. These three datasets are frequently used as public benchmarks and differ in size and density, which gives us the opportunity to test models on different kinds of datasets.
\begin{table}[t]
  \small
  \centering
  \caption{Dataset statistics after preprocessing. `\#Inter.' denotes the number of interaction. '\#Len$_\mathcal{U}$' and `\#Len$_\mathcal{I}$' denotes average actions of Users/Items.}
    \begin{tabular}{c|cccccc}
    \hline
    Datasets & \#Inter. & \multicolumn{1}{l}{\#Users} & \multicolumn{1}{l}{\#Items} & \multicolumn{1}{l}{\#Len$_\mathcal{U}$} & \multicolumn{1}{l}{\#Len$_\mathcal{I}$} & Sparsity \\
    \hline
    \hline
    Toys & 167,597& 19,412 & 11,924 & 8.6  & 14.1 &  99.93\%\\
    Sports & 296,337 &  35,598 & 18,357 & 8.3  & 16.1 & 99.95\%\\
    Beauty & 198,502 & 22,363 &  12,101 & 8.9  & 16.4 & 99.93\%\\
    \hline
    \end{tabular}
  \label{tab:tab1}
\end{table}
\begin{table*}[t]
    \caption{Comparison of recommendation performance with baseline models in numerical values. Bold scores are the best in each row, while underlined scores are the second best.
    $^*$ and $^\dagger$ denote CNN-based and Transformer-based models, respectively.
    }
    \resizebox{\linewidth}{!}{
    \begin{tabular}{c|l|ccccccccc|c|c}
    \hline
         \multirow{2}*{Dataset} & \multirow{2}*{Metric} & (a) & (b) & (c) & (d) & (e) & (f) & (g) & (h) & (i) & (j) & (k)\\
         \cline{3-13}
         &  & GRU4Rec & Caser$^*$ & NextItNet$^*$ & SASRec$^\dagger$ & BERT4Rec$^\dagger$ & HGN & GCSAN$^\dagger$ & SINE & CORE$^\dagger$ & AdaMTC & Improv.\\
         \hline
         \hline
         \multirow{5}*{Toys}& Recall@1& 0.1650 & 0.1264 & 0.1092 & \underline{0.2120} &  0.1298 & 0.1679 & 0.1840 & 0.1745 &0.1785  & \textbf{0.2184} & +3.02\% \\
         & Recall@5& 0.3535 & 0.3144 & 0.2893& \underline{0.3956} &0.3040& 0.3609 & 0.3530 & 0.3678 & 0.3812 &  \textbf{0.4132} & +4.45\%\\
         &Recall@10& 0.4591  & 0.4231 & 0.3952 & \underline{0.4886} & 0.4097 &0.4683  & 0.4469 & 0.4665 & 0.4860 & \textbf{0.5088}  & +4.13\%\\
         &NDCG@5& 0.2630 & 0.2237& 0.2019 & \underline{0.3080} & 0.2195 & 0.2679 & 0.2718 & 0.2748 & 0.2842 & \textbf{0.3203}  & +3.99\%\\
         &NDCG@10& 0.2971 & 0.2588 & 0.2361 & \underline{0.3380} & 0.2535 & 0.3024 & 0.3021 & 0.3066 & 0.3180 & \textbf{0.3511}  & +3.88\%\\
         \hline
         \multirow{5}*{Beauty}&Recall@1&0.1848 & 0.1503 & 0.1580 & \underline{0.2154} & 0.1457 & 0.1835 & 0.1832 & 0.1903 & 0.1826 & \textbf{0.2178} & +1.11\%\\
         & Recall@5& 0.3688& 0.3364 & 0.3560 & \underline{0.4014} & 0.3176 & 0.3793 &  0.3629 & 0.3868 & 0.3949 & \textbf{0.4207}  & +4.81\%\\
         &Recall@10 &0.4687  & 0.4370 & 0.4605 & 0.4933 & 0.4163 & 0.4792 & 0.4554 & 0.4806 & \underline{0.4951} & \textbf{0.5178}  & +4.58\%\\
         &NDCD@5 & 0.2806 & 0.2469 & 0.2611 & \underline{0.3127} & 0.2348 & 0.2859 & 0.2771 &0.2937 & 0.2938 & \textbf{0.3241} & +3.65\%\\
         &NDCD@10& 0.3129 & 0.2793 & 0.2949 & \underline{0.3424} & 0.2666 & 0.3181 & 0.3069 & 0.3239 & 0.3262 & \textbf{0.3554}  & +3.80\%\\
         \hline
         \multirow{5}*{Sports}&Recall@1 & 0.1423 & 0.1239 &0.1255& \underline{0.1716} & 0.1178 & 0.1593 & 0.1369 & 0.1641 & 0.1453 & \textbf{ 0.1862}  & +8.51\%\\
         &Recall@5 & 0.3386 & 0.3173 & 0.3275 & \underline{0.3760} & 0.3036 & 0.3657 & 0.3303 & 0.3751 & 0.3713 & \textbf{0.4097}  & +8.96\%\\
         &Recall@10 & 0.4606 & 0.4417 & 0.4578 & \underline{0.4990} & 0.4269 & 0.4890 & 0.4457 & 0.4935 & 0.4933 & \textbf{0.5412}  & +8.46\%\\
         &NDCG@5 &  0.2434 & 0.2226 & 0.2287 &  \underline{0.2775} & 0.2128 & 0.2660 & 0.2365 & 0.2728 & 0.2616 & \textbf{0.3019} & +8.79\%\\
         &NDCG@10& 0.2828 & 0.2627 &0.2708  & \underline{0.3172} & 0.2525 & 0.3059& 0.2737 & 0.3110 & 0.3010 &\textbf{0.3444}  & +8.58\%\\
         \hline
    \end{tabular}}
    \label{tab:tab2}
\end{table*}
\subsubsection{Evaluation Metrics}
To evaluate the performance of each method, we employ the widely used \emph{leave-one-out} evaluation task for each user's item sequence, which means we hold out the last item for the test, the second last item for validation, and the rest for training. Besides, we follow the common practice in \citep{kang2018self,li2020time,sun2019bert4rec,cho2020meantime} to pair each ground truth item in the test set with 100 randomly sampled \emph{negative} items that have not been interacted by the user. To make the task more realistic, we sample the negative items by uniform samplings. To score the ranked list, we employ two common evaluation metrics: \emph{Hit Ratio} (HR), and \emph{Normalized Discounted Cumulative Gain} (NDCG). Since we only have one ground truth item for each user, HR@\emph{K} is equivalent to Recall@\emph{K}. Therefore, both Recall@\emph{K} and NDCG@\emph{K} have large values when the ground truth item is ranked higher up in the top-\emph{K} list. We report Recall and NDCG with $k=1,5,10$.

\subsubsection{Baseline Models}
We choose some typical state-of-the-art SR methods for comparison, especially including individual CNNs (Caser, NextItNet) and Transformer (SASRec, BERT4Rec, GCSAN, and CORE). The baselines are introduced as follows. 
\begin{itemize}
    \item \textbf{GRU4Rec} \citep{hidasi2015session}: A session-based recommender system based on recurrent neural network. 
    \item \textbf{Caser} \citep{Caser}: A CNN-based model with horizontal and vertical convolutional layers for personalized recommendation.
    \item \textbf{NextItNet}  \citep{yuan2019simple}: A simple convolutional generative network for next item recommendation.
    \item \textbf{SASRec} \citep{kang2018self}: A Transformer-based model that uses self-attention network for the sequential recommendation.
    \item \textbf{BERT4Rec} \citep{sun2019bert4rec}: A bidirectional Transformer-based model for the sequential recommendation.
    \item \textbf{HGN} \citep{ma2019hierarchical}: A hierarchical gating networks for sequential recommendation.
    \item \textbf{GCSAN} \citep{xu2019graph}: A graph contextualized self-atttention network for session-based recommendation.
    \item \textbf{SINE} \citep{tan2021sparse}: A sparse-interest network for sequential recommendation.
    \item \textbf{CORE} \citep{hou2022core}: A simple and effective session-based recommendation with unifying representation space.
\end{itemize}
\subsubsection{Implementation Details}
We evaluate baseline models based on the famous open-source RecBole framework\footnote{\url{https://github.com/RUCAIBox/RecBole}}. We apply grid search to find the optimal hyper-parameters for each model. We consider hidden dimension size $d_{model} \in \{16, 32, 64, 96, 128\}$, number of heads $h \in \{2, 4, 8, 16, 32\}$, reduction ratio $r \in \{1,2,5,10,25\}$, kernel size $k \in \{3,5,7,9,11\}$, hidden activation $\Phi \in \{`GELU',`ReLU',\\
`Swish',`TanH',`Sigmoid'\}$, hidden dropout ratio $d_1\in [1e-1, 1]$, attention dropout ratio $d_2 \in [1e-1, 1]$ and learning rate $lr \in [1e-8, 1]$. 
We report the results of each baseline under its optimal hyper-parameter settings. For our AdaMCT, the optimal hyper-parameters for each benchmark are $\textbf{Toys}:\{L=2,k=3,r=2,\Phi=`GELU',h=4,d_{model}=32,d_1=0.50,d_2=0.50,lr=0.001\}$, $\textbf{Beauty}:\{L=2, k=5,r=2,\Phi=`Swish',h=2,d_{model}=16,d_1=0.43, d_2=0.57,lr=0.02\}$, and $\textbf{Sports}:\{L=2, k=3,r=5,\Phi=`GELU',h=4,d_{model}=64,d_1=0.76,d_2=0.51,lr=0.001\}$, respectively. 
We train the model using the Adam algorithm with $\beta_1=0.9$, $\beta_2=0.999$, and $l_2$ weight decay of $0.0$, and linear decay of the learning rate. To ensure fairness, we used the same maximum sequence length ($N=50$ for Amazon) and the number of layers ($L=2$) as in \citep{kang2018self}. We train each model with early stop strategies until the validation accuracy does not improve for 20 epochs on a single NVIDIA A100 SXM4 80GB GPU with a batch size of 2048. The results of the test sets are reported in the following section.
\begin{figure*}[t]
	\centering
	\includegraphics[width=0.9\linewidth]{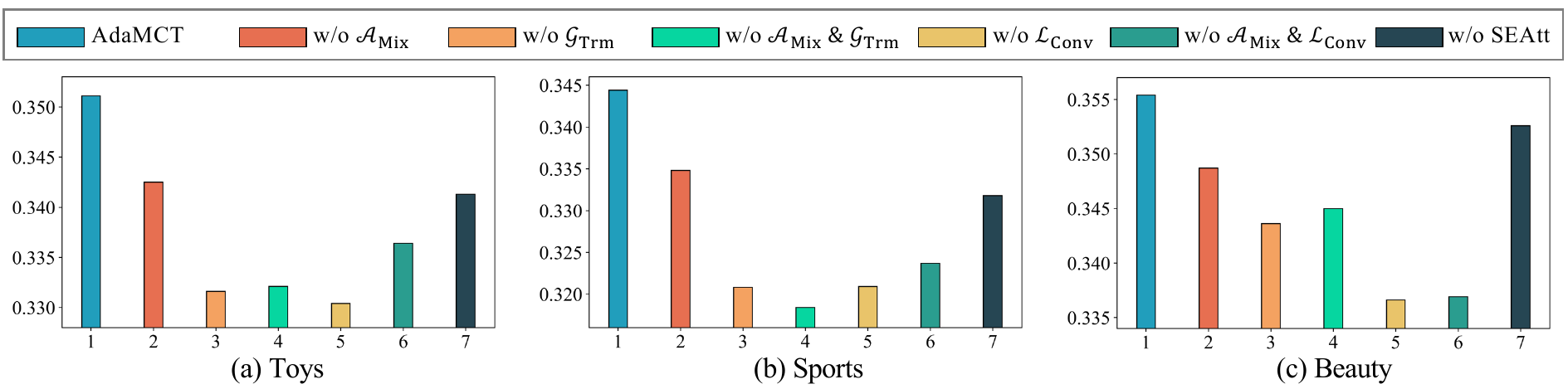}
	\caption{Performance (NDCG@10) comparison of with(w) or without(w/o) proposed components including a global attention module (Transformer) $\mathcal{G}_\text{Trm}$, a local convolutional module (CNNs) $\mathcal{L}_\text{Conv}$, an adaptive mixture unit $\mathcal{A}_\text{Mix}$, and two squeeze-excitation attention (SEAtt) per layer of AdaMCT on three datasets.
	}
	\label{fig:dim}
\end{figure*}
\subsection{Overall Performance (RQ1)}
Table~\ref{tab:tab2} shows the recommendation performance of our model and baselines on the three datasets. Here, we use the averaged performance run by five random seeds. Consequently, the improvement provided by AdaMCT is statistically and consistently significant over all datasets. 

We highlight two important baselines Caser and SASRec which belong to the individual architecture of CNNs and Transformer respectively. AdaMCT surpasses Caser by a margin exceeding 4\% on the Toys and Beauty datasets, and even over 8\% on the Amazon Sports dataset, considering all metrics on average. Caser performs fairly well compared to GRU4Rec, implying that models containing local inductive bias are nontrivial in predicting the next item. Self-attention based models such as SASRec and BERT4Rec show better performances over the RNN-based GRU4Rec and CNN-based Caser. This suggests that global attentive representation learning is relatively more important than the use of only one inductive bias. Meanwhile, GNN-based models, \textit{i.e.,} GCSAN outperforms RNN-based and CNN-based models, underscoring the importance of capturing pairwise relationships between items. Nevertheless, these models fall short of surpassing SASRec, indicating that sequential relationships prevail when modeling user preferences in sequential recommendation.

On average, AdaMCT outperforms SASRec across all three datasets. As such, AdaMCT emerges as the most effective model, adeptly considering both \emph{local} and \emph{global} dependencies of user interest patterns in an adaptive manner.

If we extend the comparison to ablated AdaMCT, we find that AdaMCT performs better than SASRec while allowing either the local module or the global module to be adaptive, as shown in Table \ref{tab:tab3}. This means that with our proposed adaptive mixture units, only CNN-based components can already be comparable with SASRec. On the contrary, when we fix the global and local trade-off to a constant value without user-specific learnable adaptability, the recommendation performance drops drastically. The best fixed AdaMCT using 80\% global and 20\% local modules without personalized adaptability is only slightly better than BERT4Rec but worse than the self-adaptive AdaMCT. Table 3 also shows that the proposed AdaMCT that adaptively exploits the mixing importance of CNNs and Transformer layer by layer can improve the overall representation learning effectively. 
\begin{table}
\centering
  \caption{Performance (NDCG@10) comparison of the adaptive or fixed mixture of local (CNNs) $\mathcal{L}_{\text{Conv}}$ and global (Transformer) $\mathcal{G}_{\text{Trm}}$ on three datasets. The best performance in each column is boldfaced.}
  \label{tab:tab3}
  \begin{tabular}{c|cc|ccc}
    \hline
    \multirow{1}*{Setting} & $\mathcal{L}_{\text{Conv}}$ & $\mathcal{G}_{\text{Trm}}$ & \multirow{1}*{Toys} & \multirow{1}*{Beauty} & \multirow{1}*{Sports} \\
    \hline
    \hline
    \multirow{4}*{Adaptive} & \xmark & \xmark & 0.3425 & 0.3487 &0.3348 \\ 
    ~ & \cmark & \xmark & 0.3429& 0.3466&0.3341 \\
    ~ & \xmark & \cmark &0.3404 & 0.3494&0.3297 \\
    ~ & \cmark & \cmark & \textbf{0.3511}& \textbf{0.3554}& \textbf{0.3444}\\
    \hline
    \multirow{5}*{Fixed} & 
        $\times 1.0$ & $\times 0.0$ &0.3322 & 0.3450 & 0.3184\\
    ~ & $\times 0.8$ & $\times 0.2$ & 0.3403& 0.3484& 0.3330\\
    ~ & $\times 0.5$ & $\times 0.5$ & 0.3412&0.3442 & 0.3314\\
    ~ & $\times 0.2$ & $\times 0.8$ & 0.345&0.3477 & 0.3334\\
    ~ & $\times 0.0$ & $\times 1.0$ & 0.3364 &0.3369 &0.3237 \\
    \hline
  \end{tabular}
\end{table}
\subsection{Ablation Study (RQ2)}
To understand the importance of different modules in our model, we conduct a detailed ablation study. We analyze the importance of self-adaptive local (CNNs) and global (Transformer) awareness in Table \ref{tab:tab3}, analyze the functions of various computational components in Figure~\ref{fig:dim}, and analyze the effectiveness of squeeze-excitation attention (SEAtt) in Table \ref{tab:tab5}.

As can be seen in Figure~\ref{fig:dim}, when we remove any of the proposed components of AdaMCT, the performance drops by a significant margin, which demonstrates the effectiveness of each designed component. Specifically, the ablation results reveal that removing the global attention module (Transformer) $\mathcal{G}_\text{Trm}$ and local convolutional module (CNNs) $\mathcal{L}_\text{Conv}$ will significantly drop the model performance across all three datasets. It confirms our intuition that local and global features are important for sequential recommendation tasks. Moreover,  adaptive mixture unit $\mathcal{A}_\text{Mix}$ plays a vital role in mixing the local (CNNs) and global (Transformer) features. Removing two squeeze-excitation attention (SEAtt) also leads to performance decline. However, different datasets have shown different sensitivity patterns. It seems that the simpler Beauty dataset is easier to be learned the importance of each item compared to the other two datasets.  
\begin{figure*}
    \centering
    \includegraphics[width=0.9\linewidth]{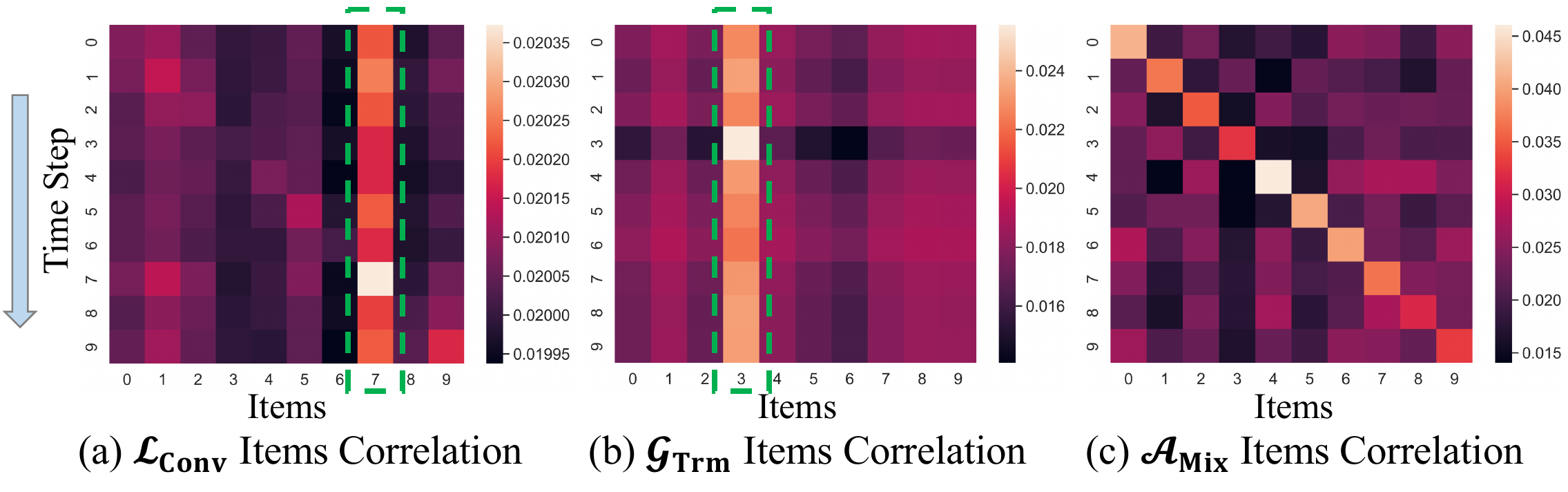}
    \caption{{Visualizations of item correlation patterns in the proposed $\mathcal{L}_{\text{Conv}}$, $\mathcal{G}_{\text{Trm}}$, and $\mathcal{A}_{\text{Mix}}$ components of AdaMCT on Beauty dataset. It is calculated by the feature representation matrics of items of each component in the last layer multiplying its transposed matrics followed by $Softmax$ normalization.}}
    \label{fig:map}
\end{figure*}
\begin{table}
\centering
  \caption{Performances (NDCG@10) comparison of Squeeze-Excitation Attention (SEAtt) with local (CNNs) $\mathcal{L}_{\text{Conv}}$ and global (Transformer) $\mathcal{G}_{\text{Trm}}$ awareness on three datasets. `$\clubsuit$' and `$\diamondsuit$' indicates adopting $Sigmoid$ and $Softmax$ activation function, respectively. The best performance in each column is boldfaced.} 
  \label{tab:tab5}
  \begin{tabular}{c|cc|ccc}
    \hline
    \multirow{1}*{Setting} & $\mathcal{L}_{\text{Conv}}$ & $\mathcal{G}_{\text{Trm}}$ & \multirow{1}*{Toys} & \multirow{1}*{Beauty} & \multirow{1}*{Sports} \\
    \hline
    \hline
    (1) & \LARGE$\diamondsuit$ & \LARGE$\diamondsuit$ &0.3343 &0.3447 & 0.3226\\
    (2) & \LARGE$\clubsuit$ & \LARGE$\diamondsuit$ & 0.3357&0.3433 & 0.3316\\
    (3) & \LARGE$\diamondsuit$ & \LARGE$\clubsuit$ & 0.3355& 0.3525& 0.3255\\
    \multirow{1}*{\makecell[c]{(4)}} & \LARGE $\clubsuit$ & \LARGE $\clubsuit$ & \textbf{0.3511}& \textbf{0.3554}& \textbf{0.3444}\\ 
    \hline
  \end{tabular}
\end{table}

We discuss the effectiveness of the squeeze-excitation attention (Sigmoid) in Table \ref{tab:tab5}. The proposed SEAtt can improve the original Softmax approach by approximately 5.03\%, 2.73\%, and 6.29\% over three datasets respectively, in terms of NDCG@10. Besides, we observe that Softmax is more beneficial to the local branches (CNNs), and Sigmoid is more favorable for the global branches (Transformer). In addition, the performance of Sigmoid on the local branch still outperforms the performance of Softmax on the global branch, and the hybrid of Softmax and Sigmoid is better than simply adopting the Softmax on both branches except the Beauty dataset. These demonstrate that considering multiple relevant items of equal importance simultaneously on user historical behavior is conducive to capturing user interest preference accurately.  
\begin{table}
\centering
  \caption{Comprehensive hyperparameter study of kernel size ($k$), reduction ratio ($r$), hidden activation ($\Phi$), head number ($h$), and hidden dimension ($d_{model}$) on Beauty datasets (more datasets results can be found in Appendix). The NDCG@10 performance is reported. The best performance is boldfaced.}
  \label{tab:tab4}
  \begin{tabular}{c|ccccc}
    \hline
    \multirow{1}*{$k$} & 3 & 5& 7& 9& 11\\
    \hline
    \hline
    & 0.3489&0.3521 &0.3541 &0.3495 &0.3528 \\
    \hline
    \multirow{1}*{$r$} & 1 & 2& 5& 10& 25\\
    \hline
    &0.3487 & 0.3526& 0.3493& 0.3452&0.3520 \\
    \hline
    \multirow{1}*{$\Phi$} & $GELU$ & $ReLU$& $Swish$& $TanH$& $Sigmoid$\\
    \hline
    &0.3461 & 0.3541& 0.3526& 0.3491&0.3496 \\
    \hline
    \multirow{1}*{$h$} & 2 & 4& 8& 16& 32\\
    \hline
    & 0.3526& 0.3470& 0.3525&0.3496 & 0.3485\\
    \hline
    \multirow{1}*{$d_{model}$} & 16 & 32& 64& 96& 128\\
    \hline
    & \textbf{0.3554}& 0.3526&0.3309 & 0.3216 & 0.3224\\
    \hline
  \end{tabular}
\end{table}
\subsection{Hyperparameter Study (RQ3)}
\begin{figure}
    \centering
    \includegraphics[width=0.7\linewidth]{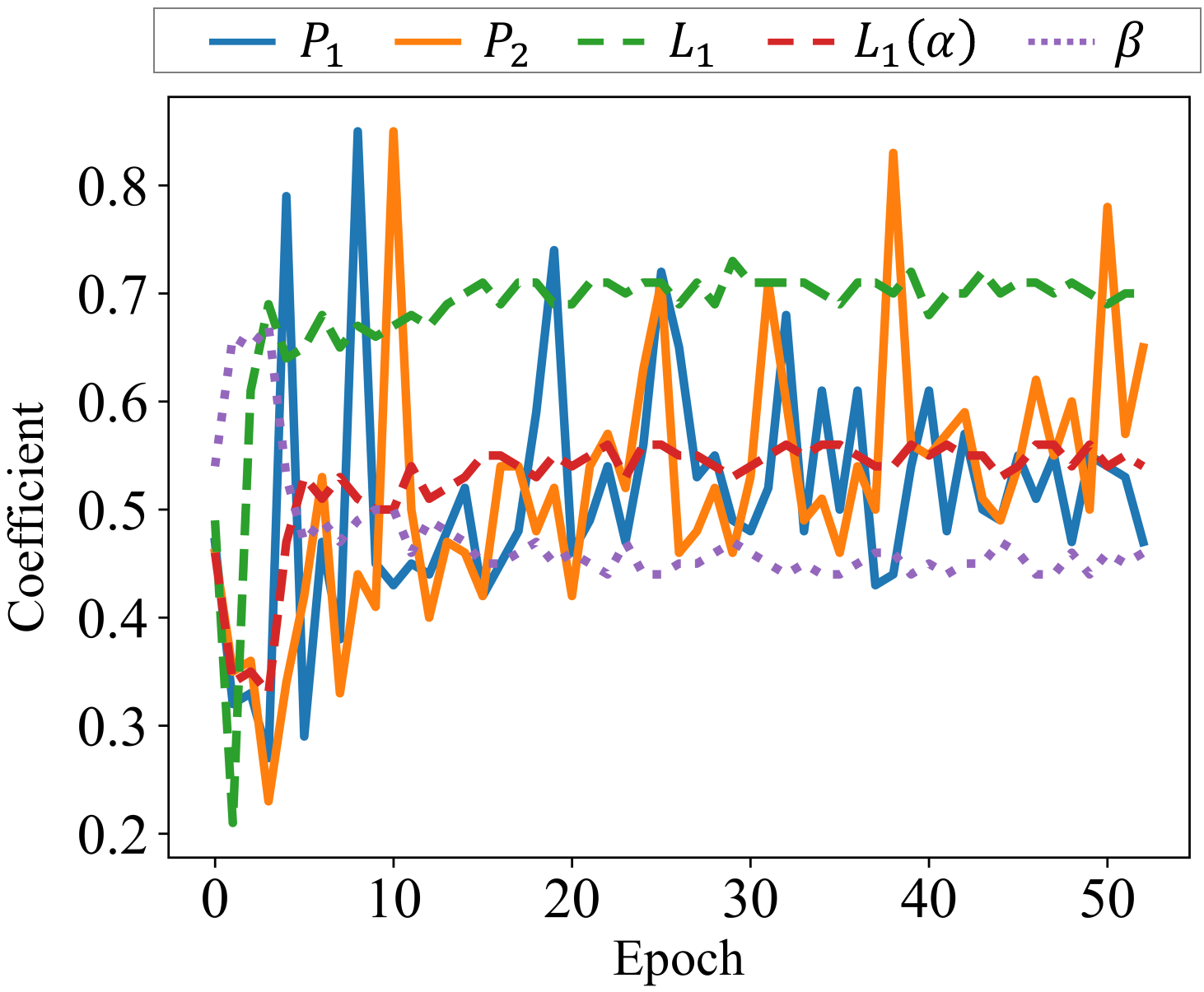}
    \caption{{Visualization of learning procedure of adaptive coefficient for two users ($P_1$ and $P_2$), two layers ($L_1$ and $L_2$), and the $\alpha$ and $\beta$ in the last layer ($L_2$) on Toys dataset during training iterations.}}
    \label{fig:coe}
\end{figure}

In this section, we conduct a stability study to discuss the effect of a variety of hyperparameters on model performance in Table \ref{tab:tab4}. We also visualize the learning procedure of local-global coefficients to get the gist of how AdaMCT adaptively mixes the feature of CNNs and Transformers, as can be seen in Figure \ref{fig:coe}. 
Specifically, we analyze the effect of the kernel size ($k$), reduction ratio ($r$), hidden activation ($\Phi$), head number ($h$), and hidden dimension ($d_{model}$) in Table \ref{tab:tab4}. It can be seen that on all three datasets, the performance of AdaMCT improves by a narrow margin with changes in dimensionality. This shows that our model is generally insensitive to dimensionality.
In particular, the smaller size of the hidden dimension ($d_{model}$) shows a better performance,
indicating that for simpler datasets (\textit{e.g.}, Beauty) the hidden dimensionality does not need to be too large. From the results, we can tell that AdaMCT is overall stable. 
\begin{table}[t]
    \caption{Parameters number and execution efficiency analysis of models on Beauty dataset.} 
    \label{tab:complexity}
    \centering
    \resizebox{\linewidth}{!}{
    \begin{tabular}{l|c|c|c|c}
    \hline
    Beauty & AdaMCT & Caser & SASRec & CORE  \\ 
    \hline
    \hline
     \# Parameters & 0.21M & 2.91M & 0.88M & 0.88M\\ 
     \# GPU Memory & 0.79GB & 0.62GB & 2.31GB  &  2.33GB\\
     \# FLOPs ($\times10^6$) & 0.34 & 11.38 & 4.98 & 4.99\\ 
     Training Cost (per epoch) & 4.75s & 34.02s & 3.65s & 3.97s\\
     Inference Cost (per epoch)  & 26.95s & 157.07s & 22.48s & 22.57s \\
     \hline
     NDCG@10 & \textbf{0.3554}  & 0.2793  & 0.3424 & 0.3262 \\
     \hline
    \end{tabular}
    }
\end{table}

One of the novelties in the proposed AdaMCT model lies in the fusion of local (CNNs) and global (Transformer) information through a self-adaptive approach.
 To validate our intuition, we visualize the learning procedure of coefficients of the local and global branches for two randomly chosen users (denoted as user1 ($P_1$) and user2 ($P_2$)), two layers (denoted as $L_1$ and $L_2$) and the $\alpha$ and $\beta$ of the last layer with average scores, as shown in Figure \ref{fig:coe}. We observe that AdaMCT updates the coefficients through learning iterations. 
 The fluctuating curves indicate that the local and global information serve distinct functions at varying training phases:  the CNNs components are given higher weights at the beginning and then AdaMCT gradually learns to assign weights to the Transformer module. This suggests that capturing local dependency may hold greater significance than the Transformer during the initial stages, and the convolutional components could potentially furnish valuable information to guide the optimization of the Transformer, thereby enhancing performance.
 Furthermore, the coefficients exhibit variation across different users, signifying that the model self-adaptively learns distinct patterns for individualized users.
This further illustrates that it is not an ideal way to set a static and equal trade-off between local and global branches, as many works \citep{hu2018local, xu2019recurrent, chen2019session, xu2020joint, he2020consistency, liu2020deep, yu2020tagnn} did in the past. In our adaptive paradigm, the model can fuse information with the dynamical use of locality bias (CNNs) to improve the performance of the Transformer. 

\subsection{Complexity and Efficiency Analysis(RQ4)}
In order to understand whether the performance improvement is attributable to the model architecture design or the escalation in the quantity of parameters, we conduct an analysis on AdaMCT's complexity and execution efficacy, as delineated in Table~\ref{tab:complexity}. Comparing with CNN-based and Transformer-based baselines, we have the following observations: (1) AdaMCT has a relatively fewer parameters. This is reasonable because our Transformer removes Feed-forward Network (FFN) and becomes more lightweight. (2) AdaMCT has better efficiency in both training (86\% less time) and inference (83\% less time) compared with Caser. Meanwhile, AdaMCT achieves comparable training/inference time and much smaller memory footprint (66\% less) and FLOPs (93\% less) with SASRec and CORE. Moreover, our model has the minimum FLOPs compared with all baselines which further enhances the feasibility of practical deployment, especially for low-resource constraints devices. The results indicate that AdaMCT is capable of achieving a good balance between computational cost and recommendation accuracy.

\subsection{Visualization (RQ5)}
To provide insights into the AdaMCT's interpretability, we visualize the item correlation matrices based on feature representation of $\mathcal{L}_{\text{Conv}}$, $\mathcal{G}_{\text{Trm}}$, and $\mathcal{A}_{\text{Mix}}$ output in the last layer of AdaMCT, using test samples from the Beauty dataset, as shown in Figure \ref{fig:map}. 
Based on our observations, we find that
different components have different item correlation patterns. $\mathcal{L}_{\text{Conv}}$ tends to learn local item dependencies which concentrate more on recent items (red-orange area with higher correlation scores), while $\mathcal{G}_{\text{Trm}}$ prefers to learn items dependencies globally, \textit{e.g.}, the 3rd item has the higher correlation with other items, as can be seen in Figure \ref{fig:map}(a) and (b) green dash rectangular box, respectively.
 
\section{Conclusion}
In this paper, we 
design a novel Adaptive Mixture of CNN-Transformer (AdaMCT) that injects locality inductive bias into Transformer by combining its global attention mechanism with a local convolutional filter, and adaptively determines the mixing importance on a personalized basis through a module and layer-aware adaptive mixture units.
Moreover, we propose Squeeze-Excitation Attention (SEAtt) to replace the mutually exclusive activation to improve the learning of sequence dependencies with multiple relevant items. 
Extensive experiments on three public real-world datasets demonstrate the effectiveness and efficiency of our proposal. 
For future studies, we will explore how to design more powerful adaptive mixture units in-depth and incorporate item-related (\emph{e.g.}, price, date of production, producer) or behavior-related (\emph{e.g.}, purchase, rate) side information into AdaMCT.
\section{Acknowledgments}
This work was partially supported by the National Natural Science Foundation of China (No. 62276099). We appreciate the NVIDIA A100 SXM4 80GB GPUs support of Upstage.

\appendix
\section{Appendices}
\subsection{Hyperparameter Study (RQ3)}
\begin{table}[h]
\centering
  \caption{Comprehensive hyperparameter study of kernel size ($k$), reduction ratio ($r$), hidden activation ($\Phi$), head number ($h$), and hidden dimension ($d_{model}$) on Toys dataset. The NDCG@10 performance is reported. The best performance is boldfaced.}
  \label{tab:hyper_toys}
  \begin{tabular}{c|ccccc}
    \hline
    \multirow{1}*{$k$} & 3 & 5& 7& 9& 11\\
    \hline
    \hline
    & \textbf{0.3511}& 0.3440&  0.3423& 0.3491& 0.3414\\
    \hline
    \multirow{1}*{$r$} & 1 & 2& 5& 10& 25\\
    \hline
    & 0.3427& 0.3406& 0.3388& 0.3392& 0.3436\\
    \hline
    \multirow{1}*{$\Phi$} & $GELU$ & $ReLU$& $Swish$& $TanH$& $Sigmoid$\\
    \hline
    & 0.3406& 0.3397& 0.3443&  0.3425& 0.3445\\
    \hline
    \multirow{1}*{$h$} & 2 & 4& 8& 16& 32\\
    \hline
    & 0.3440& 0.3406& 0.3374& 0.3388& 0.3323\\
    \hline
    \multirow{1}*{$d_{model}$} & 16 & 32& 64& 96& 128\\
    \hline
    & 0.3454& 0.3406& 0.3413& 0.3391 & 0.3369\\
    \hline
  \end{tabular}
\end{table}
\begin{table}[h]
\centering
  \caption{Comprehensive hyperparameter study of kernel size ($k$), reduction ratio ($r$), hidden activation ($\Phi$), head number ($h$), and hidden dimension ($d_{model}$) on Sports dataset. The NDCG@10 performance is reported. The best performance is boldfaced.}
  \label{tab:hyper_sports}
  \begin{tabular}{c|ccccc}
    \hline
    \multirow{1}*{$k$} & 3 & 5& 7& 9& 11\\
    \hline
    \hline
    & 0.3383& 0.3429& 0.3424& 0.3403&0.3435 \\
    \hline
    \multirow{1}*{$r$} & 1 & 2& 5& 10& 25\\
    \hline
    & 0.3380& 0.3388& 0.3383& 0.3398& 0.3334\\
    \hline
    \multirow{1}*{$\Phi$} & $GELU$ & $ReLU$& $Swish$& $TanH$& $Sigmoid$\\
    \hline
    & 0.3383& 0.3349& 0.3347& 0.339& 0.3350\\
    \hline
    \multirow{1}*{$h$} & 2 & 4& 8& 16& 32\\
    \hline
    & 0.3383 &\textbf{0.3444} &0.3392 &0.3370 & 0.3361\\
    \hline
    \multirow{1}*{$d_{model}$} & 16 & 32& 64& 96& 128\\
    \hline
    &0.3081 &0.3370 & 0.3383 & 0.3373&0.3381 \\
    \hline
  \end{tabular}
\end{table}
\subsection{Visualization (RQ5)}
\begin{figure*}
    \centering
    \includegraphics[width=0.9\linewidth]{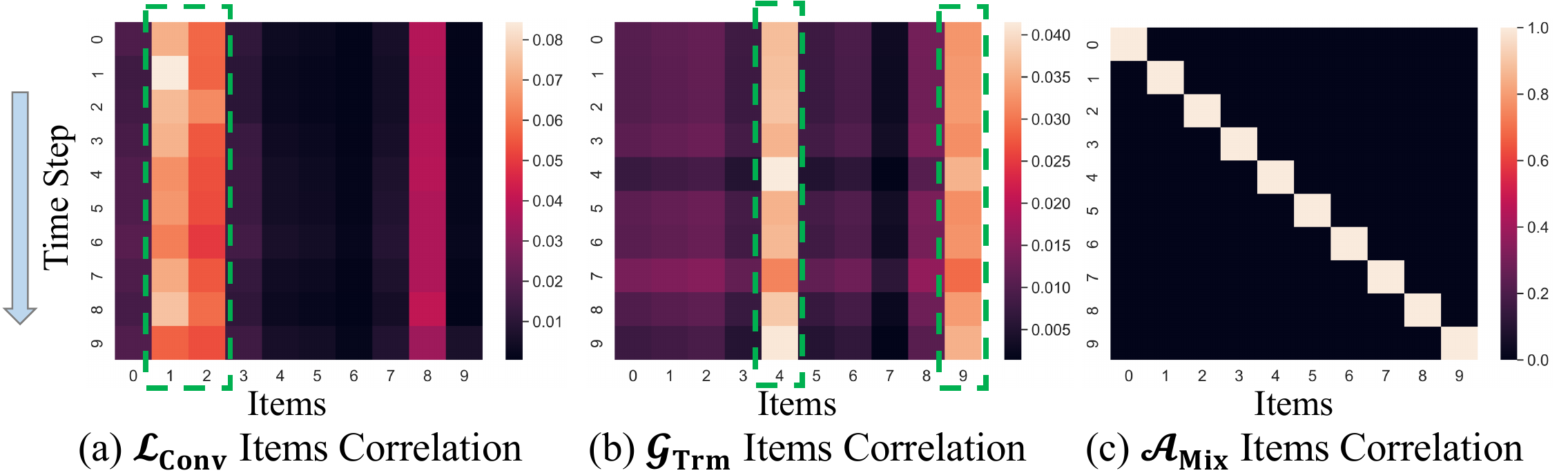}
    \caption{{Visualizations of item correlation patterns in the proposed $\mathcal{L}_{\text{Conv}}$, $\mathcal{G}_{\text{Trm}}$, and $\mathcal{A}_{\text{Mix}}$ components of AdaMCT on Toys dataset. It is calculated by the feature representation matrics of items of each component in the last layer multiplying its transposed matrics followed by $Softmax$ normalization.}}
    \label{fig:map_toys_2}
\end{figure*}
\begin{figure*}
    \centering
    \includegraphics[width=0.9\linewidth]{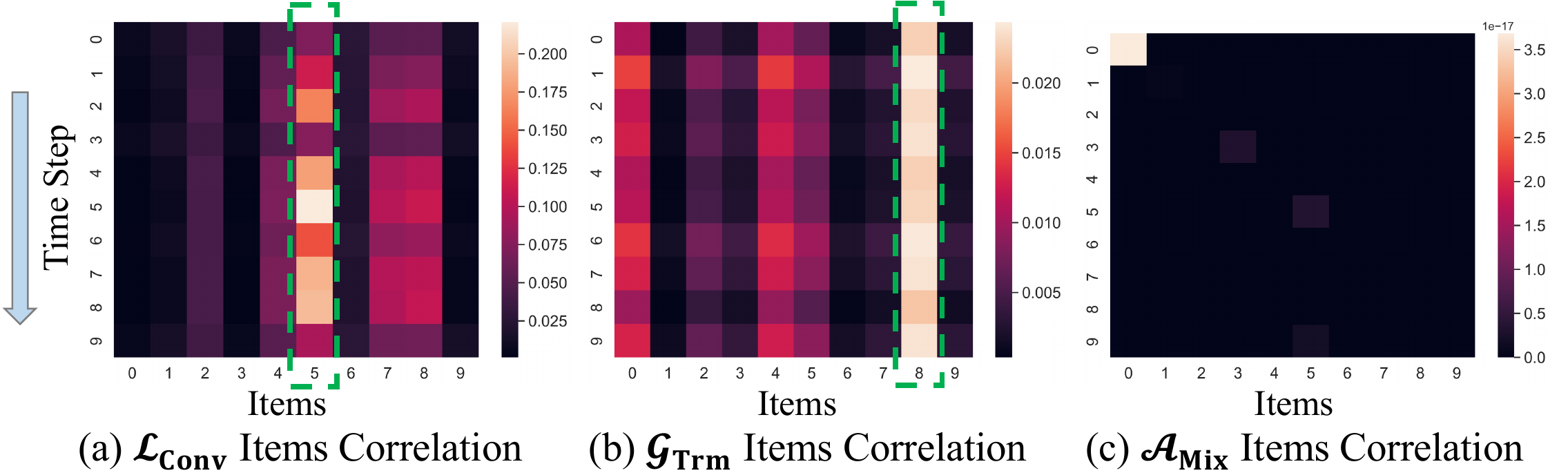}
    \caption{{Visualizations of item correlation patterns in the proposed $\mathcal{L}_{\text{Conv}}$, $\mathcal{G}_{\text{Trm}}$, and $\mathcal{A}_{\text{Mix}}$ components of AdaMCT on Sports dataset. It is calculated by the feature representation matrics of items of each component in the last layer multiplying its transposed matrics followed by $Softmax$ normalization.}}
    \label{fig:map_sports}
\end{figure*}
\begin{table}[h]
    \caption{Parameters number and execution efficiency analysis of models on Toys dataset.} 
    \label{tab:complexity_toys}
    \centering
    \resizebox{\linewidth}{!}{
    \begin{tabular}{l|c|c|c|c}
    \hline
    Toys & AdaMCT & Caser & SASRec & CORE  \\ 
    \hline
    \hline
     \# Parameters & 0.41M & 2.71M & 0.87M & 0.87M \\ 
     \# GPU Memory & 1.33GB & 0.62GB & 2.31GB & 2.33GB \\
     \# FLOPs ($\times10^6$) & 1.02 & 11.38 & 4.98 & 4.99\\ 
     Training Cost (per epoch) & 4.38s & 28.33s & 3.07s & 3.28s \\
     Inference Cost (per epoch) & 27.91s & 137.24s & 19.62s & 19.62s \\
     \hline
     NDCG@10 & \textbf{0.3511}  & 0.2588 & 0.3380 & 0.3180\\
     \hline
    \end{tabular}
    }
\end{table}
\begin{table}[h]
    \caption{Parameters number and execution efficiency analysis of models on Sports dataset.} 
    \label{tab:complexity_sports}
    \centering
    \resizebox{\linewidth}{!}{
    \begin{tabular}{l|c|c|c|c}
    \hline
    Sports & AdaMCT & Caser & SASRec & CORE  \\ 
    \hline
    \hline
     \# Parameters & 1.26M & 4.16M & 1.28M & 1.28M \\ 
     \# GPU Memory & 1.94GB  & 0.66GB & 2.16GB & 2.31GB \\
     \# FLOPs ($\times10^6$) & 3.81 & 11.38 & 4.98 & 4.99 \\ 
     Training Cost (per epoch) & 7.77s & 48.96s & 4.89s & 5.82s\\
     Inference Cost (per epoch) & 55.18s & 250.69s & 32.23s & 35.53s \\
     \hline
     NDCG@10 & \textbf{0.3444}  & 0.2627 & 0.3172 & 0.3010\\
     \hline
    \end{tabular}
    }
\end{table}
\begin{figure}
	\centering
	\subfigure[Beauty]{
		\begin{minipage}[b]{0.45\linewidth}
			\includegraphics[width=1\textwidth]{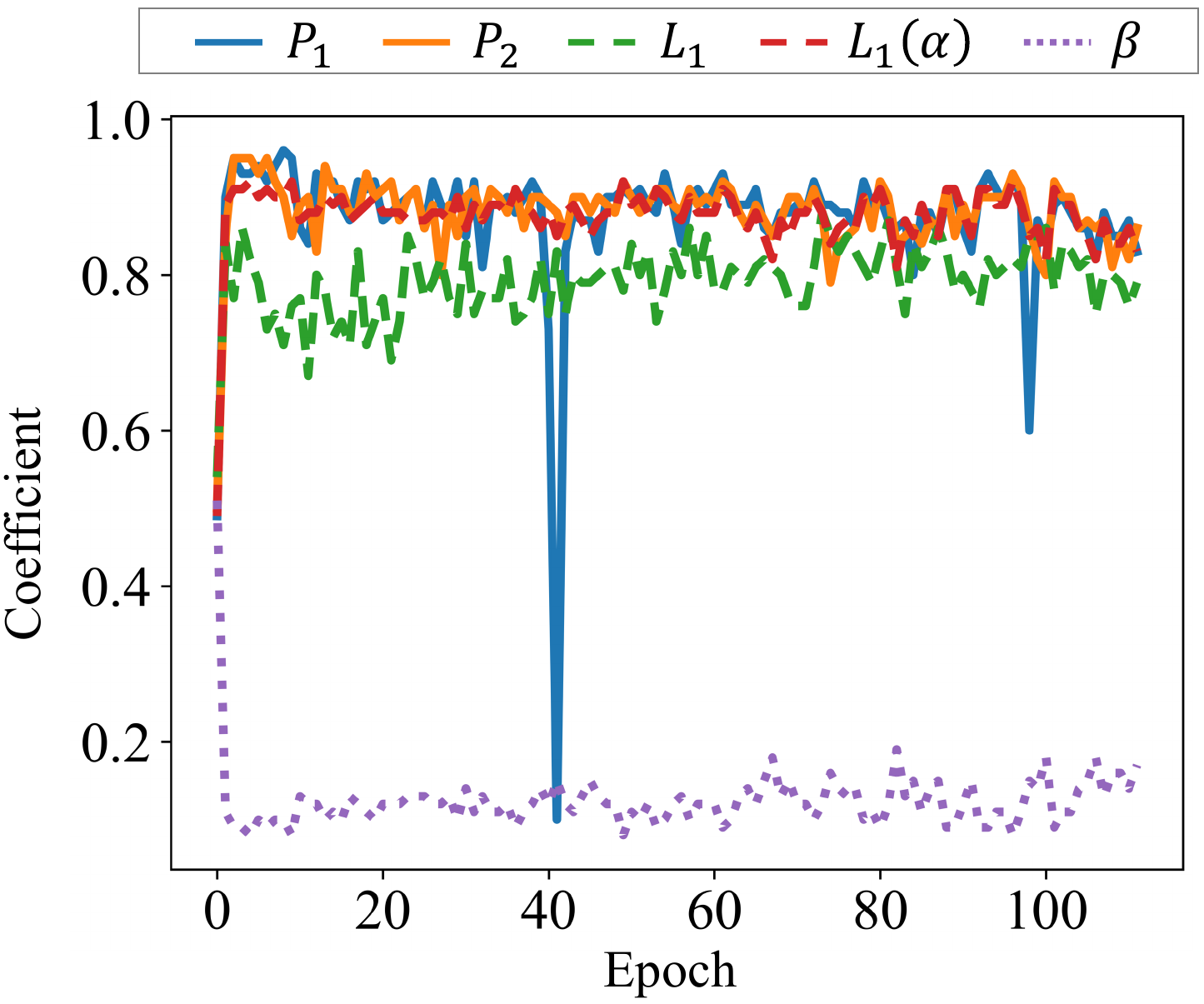}
		\end{minipage}
		\label{fig:coe_beauty}
	}
    	\subfigure[Sports]{
    		\begin{minipage}[b]{0.45\linewidth}
   		 	\includegraphics[width=1\textwidth]{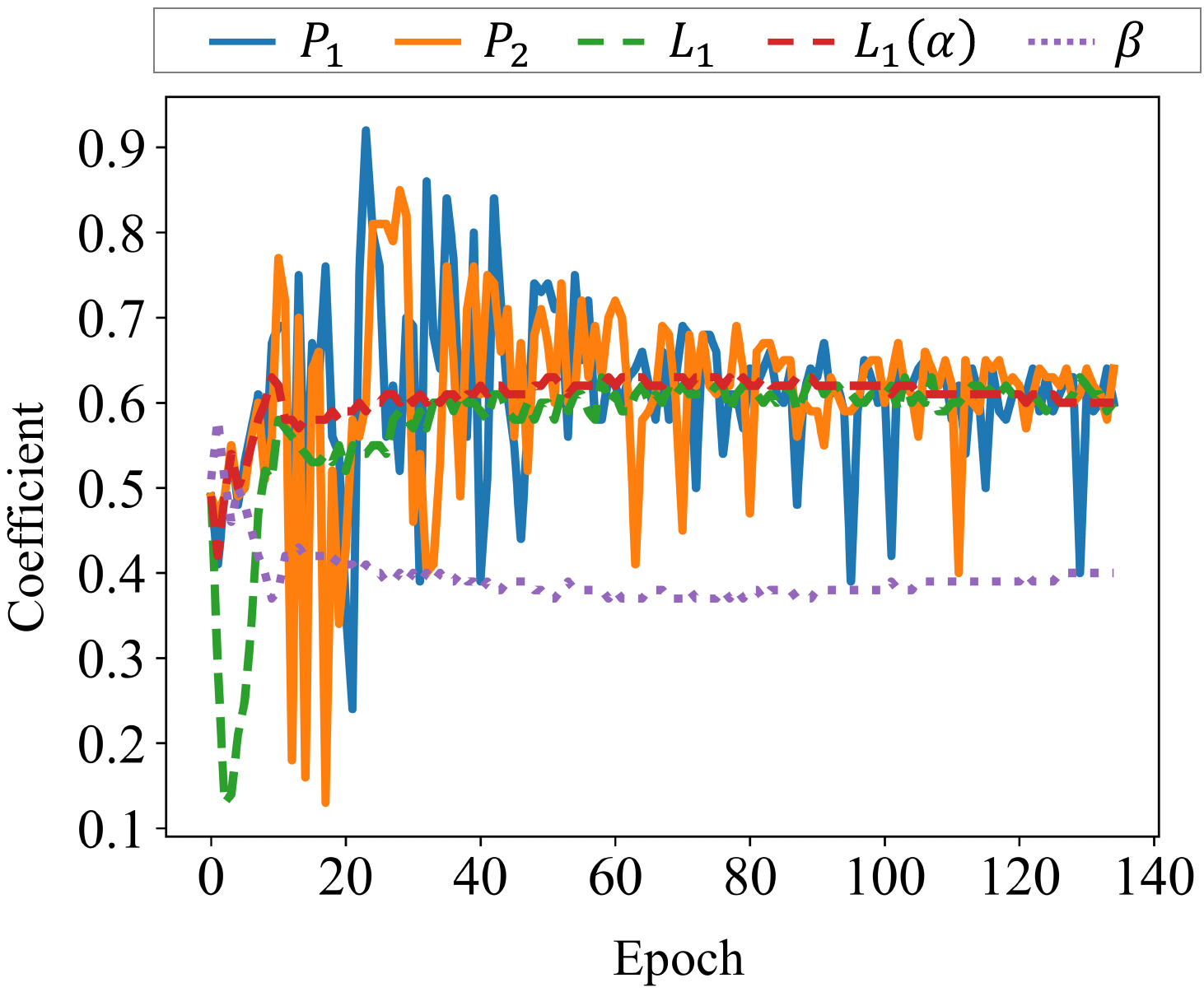}
    		\end{minipage}
		\label{fig:coe_sports}
    	}
	\caption{Visualization of learning procedure of adaptive coefficient for two users ($P_1$ and $P_2$), two layers ($L_1$ and $L_2$), and the $\alpha$ and $\beta$ in the last layer ($L_2$) on Beauty (a) and Sports (b) dataset during training iterations.}
	\label{fig:coe_beauty_sports}
\end{figure}

\clearpage
\bibliographystyle{ACM-Reference-Format}
\balance
\bibliography{ref}




\end{document}